\documentclass[preprint,superscriptaddress]{revtex4}
\usepackage{graphics,epsfig}
\usepackage{graphicx}
\usepackage{dcolumn}
\usepackage{amsmath}
\usepackage{epstopdf}
\newcommand{\be}{\begin{equation}}
\newcommand{\ee}{\end{equation}}
\newcommand{\beq}{\begin{equation}}
\newcommand{\eeq}{\end{equation}}
\newcommand{\bea}{\begin{eqnarray}}
\newcommand{\eea}{\end{eqnarray}}
\newcommand{\nn}{\nonumber}
\def\be{\begin{equation}}
\def\ee{\end{equation}}
\def\ba{\begin{eqnarray}}
\def\ea{\end{eqnarray}}

\begin{document}
%opening
\title{Thermodynamic properties of Bardeen black holes in dRGT massive gravity}
\author{R P Singh}
\email{rpsinghmathura@gmail.com}
\author{ B K Singh}
\email{bksingh100@yahoo.com}
\author{ B R K Gupta}
\email{brk.gupta@gla.ac.in}
\affiliation{Department of Physics, Institute of Applied Sciences and Humanities, GLA University, Mathura 281406, Uttar Pradesh, India}
\author{Shobhit Sachan}
\email{shobhitsachan@gmail.com}
\affiliation{Department of Physics, Sri Ram Swaroop Memorial University, Barabanki, Uttar Pradesh, India, \\
$^2$Daulat Ram College, University of Delhi, India.}

\begin{abstract}
\noindent The Bardeen black hole solution is the first spherically symmetric regular black hole based on the Sakharov and
Gliner proposal which is a modification of the Schwarzschild black hole. We present the Bardeen black hole solution in the
presence of the de Rham, Gabadaadze, and Tolly (dRGT) massive gravity, which is regular everywhere in the presence of a nonlinear
source. The obtained solution reduces to the Bardeen black hole in the absence of a massive gravity parameter, and the
Schwarzschild black hole when magnetic charge $g = 0$. We investigated the thermodynamic quantities, that is, mass (M), temperature
$(T)$, entropy $(S$), and free energy $(F)$, in terms of the horizon radius for both canonical and grand canonical ensembles. We
checked the local and global stability of the obtained solution by studying the heat capacity and free energy. The heat capacity
changes sign at $r = r_c$. The black hole is thermodynamically stable with a positive heat capacity $C > 0$ (i.e., globally preferred
with negative free energy $F < 0$). In addition, we studied the phase structure of the obtained solution in both ensembles.

\end{abstract}

\maketitle

%-----------------------------------------------------------------------
\section{Introduction}

Einstein’s general relativity is a unique theory of gravity, and
black holes are one of its exact solutions, and are characterised
by the no-hair theorem. The boundary of a black hole is known as
the event horizon, which is a one-way surface, meaning that
nothing can escape from it, including electromagnetic radiation.
The existence of a singularity means that space–time ceases to
exist, signalling the breakdown of general relativity, requiring
modifications in said theory. Sakharov \cite{Sakharov:1966}  and Gliner \cite{Gliner:1966} proposed
a method to resolve the singularity problem by considering a de
Sitter core with equation of state $P=-\rho$ or to obtain a regular
model without singularities. This model could provide proper
discrimination at the final stage of gravitational collapse, replacing
the future singularity. Using this idea, Bardeen  \cite{Regular} gave the
first black hole solution that shows that there are horizons, but
there is no singularity. These solutions are an exact solution
of general relativity coupled with nonlinear electrodynamics
(NLED) proposed by Ayon-Beato and Garcia  \cite{AGB,AGB1,ABG99}. Subsequently,
significant efforts have been made to investigate regular black
holes \cite{Ansoldi:2008jw,Lemos:2011dq,Zaslavskii:2009kp,Bronnikov:2000vy} (more recently refs. \cite{hc,lbev,Balart:2014cga,Xiang,singh,kumar:2019wpu,Singh:2019wpu,Kumar:2020bqf,dvs99,Tzikas:2018cvs,Singh:2020xju}), but most of these solutions
are fundamentally based on Bardeen’s proposal. Regular
black holes are also found in Einstein–Gauss–Bonnet gravity \cite{25,28,29,Singh20}, $f(r)$ gravity \cite{33}, quadratic gravity \cite{34}, $f(T)$ gravity  \cite{35}, noncommutative
geometry  \cite{27}, rotating black hole solution \cite{31,32},
and P-V criticality \cite{s1,s2,s3,s4,s5}.

The Einstein–Hilbert action in space–time coupled with NLED
is expressed as \cite{AGB1,dvs99},
\be
I =\frac{1}{2 }\int d^{4}x\sqrt{-g}\Big[ \mathcal{R} +{\cal{L}}(F)\Big],
\label{action1}
\ee
 where ${\cal R}$ is the Ricci scalar and ${\cal{L}}(F)$ is the Lagrangian density of the nonlinear field which is given by \cite{AGB1,Singh:2020xju, dvs99}
\be
{\cal{L}}(F)=\frac{3}{2sg^2}\left(\frac{\sqrt{2g^2F}}{1+\sqrt{2g^2F}}\right)^{5/2},
\ee
where $F$ is a function of $F_{ab}F^{ab}$,  $F_{ab}$ is the electromagnetic field tensor and $s$ is the parameter which is related to the mass and charge via $s=g/2M$. For spherically symmetric space-times, the  only non-vanishing component of $F_{ab}$ is $F_{\theta\phi}$.

 Variation of the action in Eq. (\ref{action1}) with respect to the metric tensor $g_{ab}$ and the electromagnetic potential $A_a$leads to
\begin{eqnarray}
&&R_{a b}-\frac{1}{2}g_{a b}R+\Lambda g_{a b}=T_{a b}\equiv2\left[\frac{\partial {\cal{L(F)}}}{\partial F}F_{a c}F_{b}^{c}-g_{a b}{\cal{L(F)}}\right],\\&&
 \nabla_{a}\left(\frac{\partial {\cal{L(F)}}}{\partial F}F^{a b}\right)=0,\qquad\qquad\qquad \nabla_{\mu}(* F^{ab})=0,
\label{Field equation}
\end{eqnarray}
The spherically symmetric black hole admits the following black hole solution
\begin{equation}
ds^2=-\left(1-\frac{2M r^2}{(r^2+g^2)^{3/2}}\right)dt^2 +\frac{1}{\left(1-\frac{2M r^2}{(r^2+g^2)^{3/2}}\right)}dr^2+r^2d\Omega_2^2,
\label{m1}
\end{equation} 
where $d\Omega_2^2=d\theta^2+ r^2\sin^2\theta$ denotes the metric on a $2D$  sphere,   $g$ is a magnetic  charge and  $M$ is the integration constant which is related to the black hole mass. The solution becomes  Schwarzschild black hole solution in the absence of a  magnetic  charge.

The General relativity is generalized into a more effective theory which can provide mass and is known  as  dRGT massive gravity. It was  introduced in de Rham, Gabadaadze and Tolly  model \cite{drgt,1,2,3,4}, which added a potential contribution to the Einstein-Hilbert action. The dRGT massive gravity is formulated such that  the equation of motion does not contain a higher derivative term, consequently the ghost field  vanishes. But the formulation of exact solutions in this theory is arduous due to the nonlinear term which leads to intricacy in calculations. Nevertheless considerable efforts have been made to procure spherically symmetric black holes  in distinct massive gravity \cite{10,11,12,14,17,18,19,20,21,22,23,24,sgg}.  Modification of  the dRGT model is based on the definition of the reference metric and the most successful reference metric was suggested by Vegh \cite{vegh}. 

  It is believed that dRGT massive gravity may provide a possible explanation for the accelerated expansion of the universe that does not require cosmological constant and has received significant attention including searches for black holes. Motivated by the work of Sakhrov, Gliner and Bardeen we present an exact black hole solution in the presence of dRGT massive gravity coupled to NLED\cite{AGB,AGB1,ABG99}.  The NLED theory is richer than the  Maxwell theory, and in the weak-field limit it reduces to Maxwell electrodynamics.  It was shown that coupling  gravity  to NLED  can remove black hole singularities.   The obtained solution is regular everywhere including $r\to 0$ and it reduces to the Schwarzschild massive black hole in the absence of magnetic charge. We  studied the thermodynamics of  black hole including the phase transition and  the effect of massive gravity parameter in it. We also studied the thermodynamic behaviour  including phase transition by observing the nature of  free energy in canonical and grand canonical ensembles.

The remainder  of this  paper is organised as follows: The  Bardeen black hole solution in  dRGT massive gravity is obtained in Section 2.  This section also contains the relevant equations of Einstein  theory coupled with NLED. The structure and location of the horizons of the Bardeen massive black holes are investigated in  Section 3.  Section 4 is devoted to the study of the thermodynamic properties of  Bardeen massive black holes. We adopt the signature ($\--$, +,+,+) for metric and use the units $8\pi G = c = 1$.
%-----------------------------------------------------------------------
\section{Bardeen black holes solution in  massive gravity}
%-----------------------------------------------------------------------T
The Einstein-Hilbert action in the presence of the cosmological constant coupled with the dRGT massive gravity and NLED is given  by
 \be
 S=\int d^4x\sqrt{-g}\left[R+{\cal{L}}(F)+m^2_g\,\mathcal{U}(g,\phi^a)\right],
\label{action}
 \ee
where $R$ , $\mathcal{U}$ and $\phi ^a $  are the  Ricci scalar,  potential for the graviton  and the  St\"uckelberg scalar respectively.   The potential $\mathcal{U}$  modifies the gravitational field with the variation of graviton mass $m_g$.  The effective potential $\mathcal{U}$ in four-dimensional spacetime is given as 
\be
\mathcal{U}(g,\phi^a)=\mathcal{U}_2+\alpha_3\mathcal{U}_3+\alpha_4\mathcal{U}_4
\label{pot}
\ee
here $\alpha_3$ and $\alpha_4$ are dimensionless free parameters  \cite{drgt,1,2,3,sgg}
\bea
&&\mathcal{U}_2\equiv [\mathcal{K}]^2-[\mathcal{K}^2]\\&&\mathcal{U}_3\equiv [\mathcal{K}]^3-[\mathcal{K}][\mathcal{K}^2]+2[\mathcal{K}^3]\\&&\mathcal{U}_4\equiv [\mathcal{K}]^4-6[\mathcal{K}]^2[\mathcal{K}^2]+8[\mathcal{K}][\mathcal{K}^3]+3[\mathcal{K}^2]^2-6[\mathcal{K}^4]
\eea
where $
\mathcal{K}_{b}^{a}=\delta_{b}^{a}-\sqrt{g^{a\sigma}f_{ab}\partial_\sigma\phi^a\partial_b\phi^b}$, $f_{ab}$ is a reference metric and square brackets represent the traces, i.e., $[\mathcal  {K}]=\mathcal {K}_{a}^{a}$ and $[\mathcal {K}^{n}]= (\mathcal K^{n})_{a}^{a}$.  The St\"uckelberg scalars $\phi^a$ are the four scalar fields which are initiated to  restore general covariance of the theory. It is observed that the interacting terms are symmetric polynomials of ${\cal K}$. The equation of motion  does not contain  higher order derivative term because of the chosen possible coefficients. We use unitary gauge $\phi^a=x^{\mu}\delta^a_{\mu}$ \cite{vegh}. In the chosen gauge, the tensor observable metric describes the five degrees of freedom of the massive graviton.  It is noted that once the scalars are  fixed, the St\"uckelberg scalars transform according to the coordinate transformation. As the  unitary gauge is preferred, employing a coordinate transform it will break the gauge condition and then incite further change in the  St\"uckelberg scalars. The gravitational potential parameters $\alpha_3$ and $\alpha_4$  used in Eq. (\ref{pot}) are described as follows,
\be
\alpha_3=\frac{\alpha-1}{3},\qquad \alpha_4=\frac{\beta}{4}+\frac{1-\alpha}{12}
\ee
The equation of motion  is obtained by varying the action (\ref{action}) with respect to $g_{ab}$,\, which is given by
\begin{eqnarray}
&&R_{ab}-\frac{1}{2}g_{ab}R+m_{g}^2X_{ab}=T_{ab}\equiv 2\left[\frac{\partial {\cal{L(F)}}}{\partial F}F_{a c}F_{b}^{c}-g_{a b}{\cal{L(F)}}\right],\nonumber\\&&
 \nabla_{a}\left(\frac{\partial {\cal{L(F)}}}{\partial F}F^{a b}\right)=0,
\label{efe}
\end{eqnarray}
where  $X_{ab}$ is the energy-momentum tensor which can be determined by varying the potential ${\cal U}$ term w.r.t. $g_{ab}$ \cite{sgg}
\bea
X_{ab}=&&\mathcal{K}_{ab}-\mathcal{K}g_{ab}-\alpha\left\{\mathcal{K}_{ab}^2-\mathcal{K}\mathcal{K}_{ab}+\frac{[\mathcal{K}]^2-[\mathcal{K}^2]}{2}g_{ab}\right\}\nonumber\\&&+3\beta\Big\{\mathcal{K}_{ab}^3-\mathcal{K}\mathcal{K}_{ab}^2+\frac{1}{2}\mathcal{K}_{ab}\left\{[\mathcal{K}]^2-[\mathcal{K}^2]\right\}-\frac{1}{6}g_{ab}\left\{[\mathcal{K}]^3-3[\mathcal{K}][\mathcal{K}^2]+2[\mathcal{K}^3]\right\}\Big\}
\eea
An additional constraint besides modified Einstein equations can be implied by using the Bianchi identities $\nabla^{a}X_{ab}=0 $.  The line element of the spherically symmetric space-time is

\begin{equation}
ds^2=-f(r)dt^2 +\frac{1}{f(r)}dr^2+r^2d\Omega_2^2,
\label{m1}
\end{equation}
with the following metric ansatz \cite{sgg} 
\be
f_{ab}=diag(0,0, c^2,c^2\sin^2\theta)
\ee
where $c$ is the constant with the choice of preceding metric, the action remain finite since it only contains the non-negative power of $f_{ab}$ \cite{sgg}.

  The $(r,r)$ component of the  modified Einstein equations (\ref{efe}) is
\bea
\frac{ r f^\prime (r) +f(r)}{r^2}-\frac{1}{r^2}-m_g^2\left(\frac{\alpha(3r-c)(r-c)}{r^2}+\frac{3\beta(r-c)^2}{r^2}+\frac{3r-2c}{r}\right)=\frac{3Me^2r^2}{(r^2+e^2)^{5/2}},
\label{16}
\eea
The Equation (\ref{16}) admits the following solution for the metric function $f(r)$,
\be
f(r)=1-\frac{2M r^2}{(r^2+g^2)^{3/2}}+\frac{\Lambda}{3} r^2+\gamma r+\zeta,\label{sol1}
\ee
with
\bea
&&\Lambda=3m_g^2(1+\alpha+\beta),\\&&
\gamma=-cm_g^2{(1+2\alpha+3\beta)},\\&&
\zeta=c^2m_g^2(\alpha+3\beta).
\eea
 In the obtained  solution (\ref{sol1}), the cosmological constant $\Lambda $  occurs naturally in the theory in terms of the graviton mass $ m_g $  which serves as the cosmological constant. The  solution (\ref{16}) reduced to  Bardeen black hole solution in the absence of massive gravity parameter $(m_g=0) $ \cite{singh,dvs99,Tzikas:2018cvs} and it is reduced Schwarzschild black hole   in the absence of massive gravity parameter and magnetic charge.

\begin{figure*} [h]
	\begin{tabular}{c c c c}
		\includegraphics[width=0.75\linewidth]{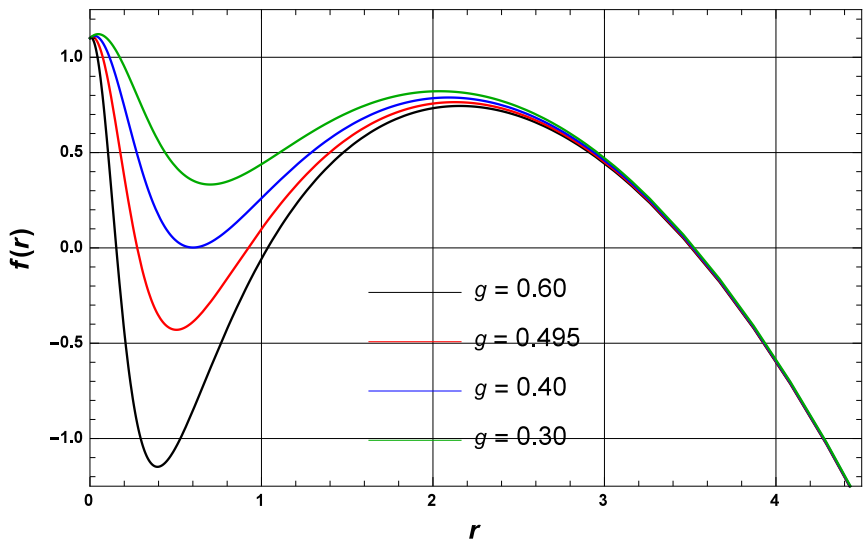}
	\end{tabular}
\caption{ Plot of  $f(r)$  vs   $r$ for different value of magnetic charge $g$ with $\alpha=2,\,\beta=0.7$ and $m_g^2c^2=1$. }
	\label{fr1}
\end{figure*}

\noindent The horizon of the black hole  can be obtained when $f(r)=0$. 
\begin{eqnarray}
1-\frac{2M r^2}{(r^2+g^2)^{3/2}}+\frac{\Lambda}{3} r^2+\gamma r+\zeta=0
\end{eqnarray}
 This is transcendental  equation it can not be solved analytically.  Numerical analysis of  $f(r)=0$ is being done by   varying the magnetic charge $(g)$ with fixed value of massive gravity parameter $m_g^2c^2=1$, is depicted in the Fig. \ref{fr1}. The numerical analysis of $f (r ) = 0$  reveals that it is possible to find non-vanishing value of $g$, $\alpha$, $\beta$ and $m_g^2c^2$   for which metric function $f (r )$ is minimum, i.e, $f (r ) = 0$, this will give three real roots which  correspond to the Cauchy horizon, event horizon and cosmological horizon. The cosmological horizon $(r_c)$ related to the graviton mass.

It is clear from Fig. \ref{fr1},  the size of the black hole increases with decrease in magnetic  charge. The black hole has three horizons at $g> 0.40$ {\it viz.} Cauchy, event, and cosmological horizon, the two horizons for $g=0.40$ viz  event and cosmological horizon and only cosmological horizon when $g>0.40$. 

Now, Let us study the nature of singularity  of   Bardeen massive black hole. It becomes useful to consider the curvature invariants of    Ricci square ($R_{ab}R^{ab}$) and Kretshmann scalars ($R_{abcd}R^{abcd}$).  The invariants are
\begin{eqnarray}
%&&\lim_{r\to 0}R=\frac{12M}{e^3}-\frac{2}{l^2},\nonumber\\
&&\lim_{r\to 0}R_{ab}R^{ab}=-12\Lambda+\frac{144M}{g^3}\left(\frac{\Lambda}{3}+\frac{M}{g^3}+\frac{ m^2 \gamma}{2} \right)+\frac{42m^2}{g^2}\left(\frac{m^2\zeta^2}{g^2}-\frac{2\Lambda\zeta}{3}+\frac{5\gamma^2}{12}+\frac{\zeta^2}{g^2} \right),\nn\\
&&\lim_{r\to 0}R_{abcd}R^{abcd}=\frac{4 \Lambda^2}{3}+\frac{48M}{g^3}\left(\frac{\Lambda}{3}+\frac{M}{g^3}+\frac{5 m^2 \zeta}{6g^2}\right)+\frac{7m^4 }{g^2}\left(\gamma^2
+\frac{6 \zeta^2}{g^2}+\frac{2g^2\zeta^2}{7}-\frac{4\zeta \Lambda}{3m^2}\right).\nn\\
\label{RR}
\end{eqnarray}
These invariants   show that the black hole  solution (17) is regular everywhere including origin ($r=0$). The singularity of the solution is removed due to the presence of Bardeen source (2).

%-----------------------------------------------------------------------
\section{Thermodynamics of the black hole}
%-----------------------------------------------------------------------   
\subsection{Canonical Ensemble}      
We investigate the thermodynamic properties of the  Bardeen  massive black hole in canonical ensemble by considering a fixed charge $g$ of the  black hole. One can determine the  mass of a black hole by $ f(r)=0 $. The mass of the black hole in terms of  horizon radius  $ r $ is given by
\bea
M=\frac{(g^2+r^2)^{3/2}\left(1+r\gamma+r^2\Lambda+\zeta\right)}{2r^2},
\eea
substituting the values of $\Lambda$, $\gamma$ and $\zeta$  from Eq. (18), (19) and Eq. (20) into Eq. (23), the mass of  Bardeen massive black hole becomes
\bea
M=\frac{(g^2+r^2)^{3/2}}{2r^2}\left(1+m_g^2r^2(1+\alpha+\beta)+c^2 m_g^2(\alpha+3\beta)-cm_g^2r(1+2\alpha+3\beta)\right).
\eea
 The mass of the  Bardeen massive black hole reduces to mass of Bardeen  black hole  in the limit of $m_g=0$ \cite{Tzikas:2018cvs} and the mass of Schwarzschild  massive black hole  when $g=0$ \cite{handi}. The black hole mass reduces to $AdS$ Schwarzschild black hole in the limit of $g=0$ and $m_g=0$.

For convenience, one can take $ m_g^2c^2=1 $, under this condition of dimensionless parameters   one can have  for positive values of black hole mass
\bea
&&\alpha>-\,\frac{r^2(1+\beta)-r(1+3\beta)+(1+3\beta)}{(r-1)^2} \qquad\text{for}\,\, r\neq 1\\&&\beta>-\,1\qquad\text{and $\alpha $ is arbitrary for}\,\, r=1.
\eea
The temperature of the black hole is known as Hawking temperature which is related to the surface gravity $\kappa$ by the relation $T=\kappa/2\pi $ \cite{Singh:2020rnm, Singh2018}. The temperature $T$ of the  black hole is
\be
T=\frac { f' (r) }{4\pi}=\frac{r^2(1+2r\gamma+r^2\Lambda+\zeta)-g^2(2+r\gamma +2\gamma)}{4\pi r(g^2+r^2)}
\label{temp1}
\ee
\begin{figure*} [h]
	\begin{tabular}{c c c c}
		\includegraphics[width=0.7\linewidth]{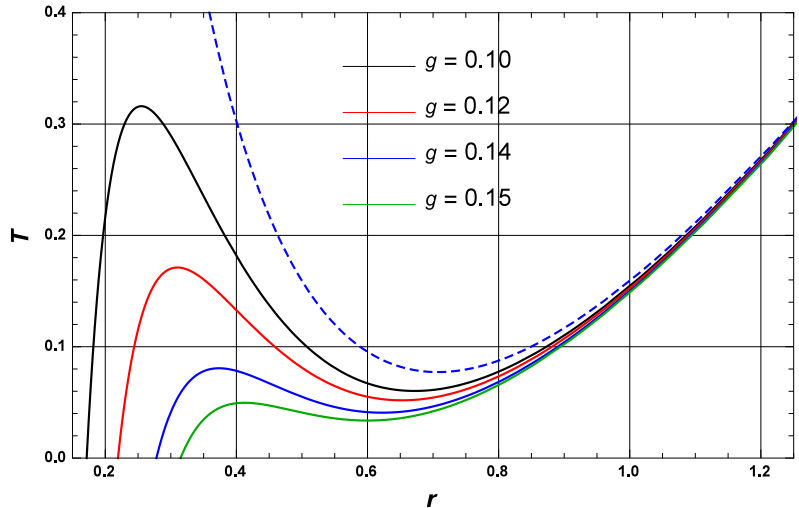}
	\end{tabular}
	\caption{ Plots of temperature of the Bardeen massive black hole $T$  vs horizon radius  $r$ for the different value of magnetic charge $g$  with $\alpha=2,\,\beta=0.7$ and $m_g^2c^2=1$. The dotted curve shows the temperature of the Schwarzschild massive black hole.}
	\label{th0}
\end{figure*}
The temperature of the Bardeen massive black hole can be recovered in the limit of graviton mass $m_g =0 $.  The plot of temperature  is displayed in Fig. \ref{th0}, which   shows that the temperature of   Bardeen massive black hole increase and attains the maximum value and then  decreases with increase in the horizon radius $r$ and attains the minimum values  $T^{min}$ for all the values of magnetic magnetic  charge which is different for the different value of magnetic e charge. After this, the temperature of the black hole increases monotonically with horizon radius and coincide with the Schwarzschild massive black hole at $r=1.1$.

Next, let us focus our attention to an important thermodynamic quantity entropy $S$ of the black hole in term of  horizon radius by using the first law of thermodynamics $dM=TdS+\phi_gdg$, where the charge is fixed. The entropy of the Bardeen massive black hole is
\be
S=\int\frac{dM}{T}=\pi r^2\left[\left(1-\frac{g^2}{r^2}\right)\sqrt{1+\frac{g^2}{r^2}}+\frac{3}{2}\frac{g^2}{r^2}\log(r+\sqrt{g^2+r^2})\right],
\ee
This entropy does not follow the area law in the presence of magnetic  charge, when $g=0$ it follow the   standard  area law which resembles with the  Bekenstein-Hawking area law.

 Wald \cite{wald} has demonstrated  that the black hole entropy  obeys the area law, but in the case of regular black holes one  does not obtain correct form of the using the first law of thermodynamics. Using the thermodynamic quantities associates with the black hole (mass, charge, temperature and entropy),  one can easily show that these quantities does not  follow the first law of thermodynamics
\be
 dM\neq TdS+\phi_g\, dg
\ee
Ma {\it et al} \cite{Ma} modified the first law  black hole thermodynamics. When the black hole mass parameter $M$ is included in the energy momentum tensor, the conventional form of the  first law gets modified with an extra factor. The corrected temperature is obtained from the modified first law of thermodynamics \cite{ Ma, Maulif,Singh:2021rnm}
\be
C_{M}dM=T\,dS  + \phi_g\,dg,
\label{mod}
\ee
 The thermodynamics variables appearing in the above expression are given by,
\bea
%&&S=\pi r^2_+,\qquad V=\frac{4}{3}\pi r^3_+ \qquad P=\frac{3}{8\pi l^2}  \nonumber \\ 
&&\phi_{g}=\frac{(g^2+r^2)^{3/2}(3+3r\gamma+r^2\Lambda+3\zeta)}{6r^2}\nonumber\\
&&C_{M} ={4\pi}\int r^2 \frac{\partial T_0^0}{\partial M}=1-\frac{r^3}{(r^2+g^2)^{3/2}}.
\label{thermo}
\eea
The entropy of the black hole
\be
S=\int C_M \frac{dM}{T}=\pi r^2=\frac{A}{4}.		
\ee

\begin{figure*} [h]
	\begin{tabular}{c c c c}
		\includegraphics[width=0.75\linewidth]{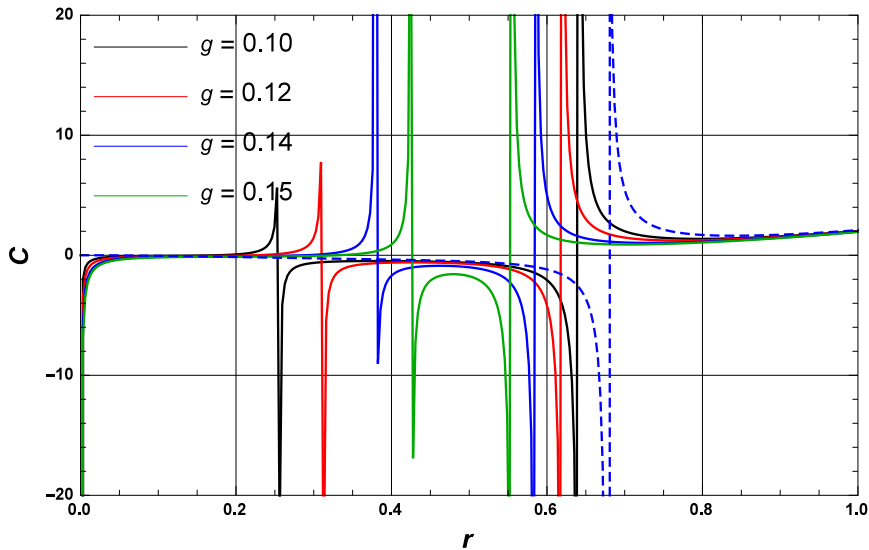}
	\end{tabular}
	\caption{ Plots of specific heat $(C)$  vs horizon radius  $r$ of the black hole for the different value of magnetic charge $g$  with $\alpha=2,\,\beta=0.7$ and $m_g^2c^2=1$. The dotted curve shows the free energy of Schwarzschild massive black hole.}
	\label{sh1}
\end{figure*}

One can also determine local thermodynamically stable state by verifying the sign of heat capacity. The heat capacity of the black hole at constant volume is defined as  
\be
C=T\left(\frac{\partial S}{\partial T}\right)
\label{hc1}
\ee
Substituting the expression of temperature and entropy in Eq.  (\ref{hc1}) and on solving it the expression for the heat capacity becomes
\be
C=\frac{2\pi(g^2+r^2)^{5/2}\left(r^2(1+2r_+\gamma+r^2\Lambda+\zeta)-g^2(2+r\gamma+2\zeta)\right)}{r\left(r^4(-1+r^2\Lambda-\zeta)+2g^4(1+c_1)+g^2r^2(7+6r\gamma+3r^2\Lambda+7\zeta) \right)}.
\ee
To analyse it,  we plot the  heat capacity in Fig. \ref{sh1} for different values of magnetic  charge   which clearly exhibits that the heat capacity for a given value of magnetic charge is discontinuous exactly at the critical radius $r_{c1}$ and $r_{c2}$. Further, it is noticeable that the black hole is   thermodynamically stable for  $r<r_{c1}$ and $r>r_{c2}$ whereas it is  thermodynamical unstable for $r_{c1}<r<r_{c2}$.  Moreover, divergence of the  heat capacity   at critical radius $r=r_{c1}$ and $r_{c2}$ indicates the occurrence of  a  phase transition   \cite{handi}. The heat capacity is discontinuous at $r_{c1}=0.252$  and $r_{c2}=0.68$ for $g=0.10$. 

\begin{figure*} [h]
	\begin{tabular}{c c c c}
		\includegraphics[width=0.5\linewidth]{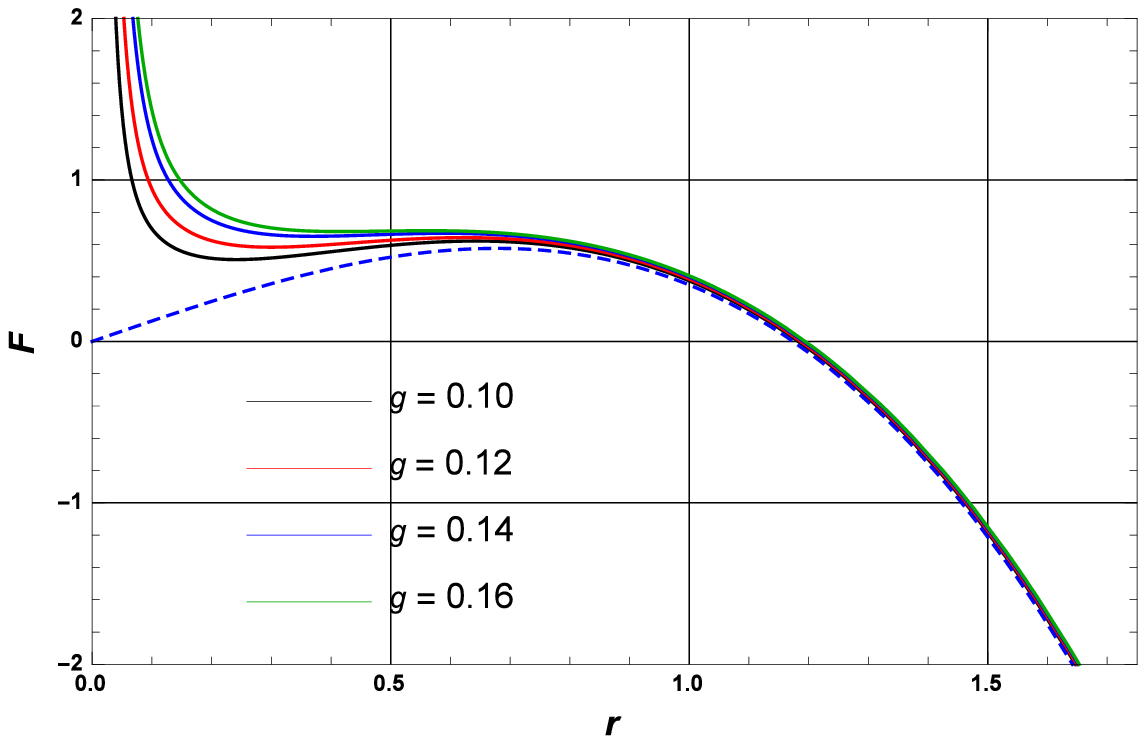}
\includegraphics[width=0.5\linewidth]{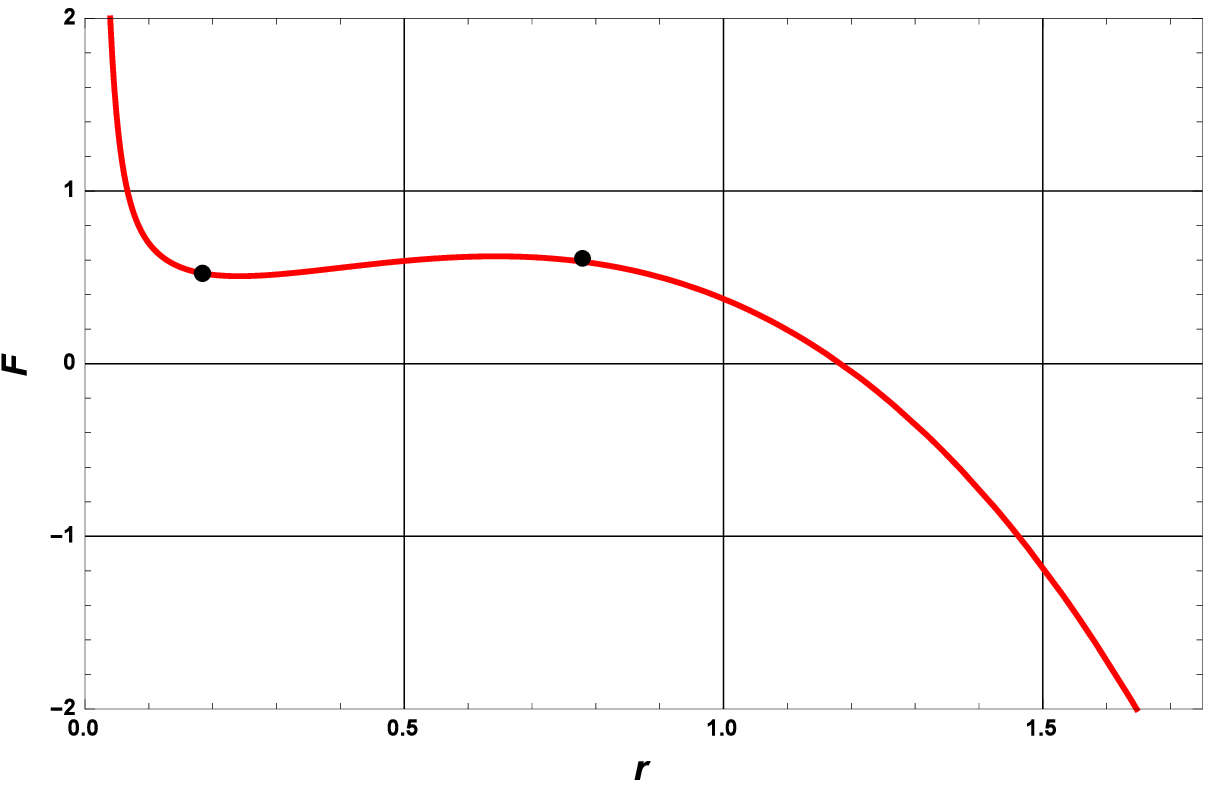}
	\end{tabular}
	\caption{ Plots of free energy $(F)$  vs horizon radius  $r$ of the black hole for the different value of magnetic charge $g$  with $\alpha=2,\,\beta=0.7$ and $m_g^2c^2=1$, and the point shows that the local minima and maxima. The dotted curve shows the Helmholtz free energy of Schwarzschild massive black hole and second plot is for $g=0.10$.}
	\label{sh20}
\end{figure*}

\noindent In  case of canonical ensemble where there is no exchange of particles and the charge is fixed, one can consider the black hole to be a closed system. For this, we turn to calculate the Helmholtz free energy  \cite{Singh:2020rnm,sgg}
\be F =M-TS
\label{feh}\ee .  Substituting the expression of  $M, T$ and $S$ in  Eq. (\ref{feh}) the Helmholtz free energy  of the Bardeen massive black hole becomes 
\bea
F&=&\frac{(g^2+r^2)^{3/2}(3+3r\gamma+r^2\Lambda+3\zeta)}{6r^2}-\frac{r\left(r^2(1+2r\gamma+r^2\Lambda+\zeta)\right)}{4(g^2+r^2)}\nn\\&&-\frac{r\left(-g^2(2+r\gamma+2\zeta)\right)}{4(g^2+r^2)},
\eea
The condition of globally thermodynamically stable black hole is give by $ F\leq 0 $. We analyse the stability of the black hole by studying the nature of free energy which is plotted in Fig.  \ref{sh20}, for the 
different values of magnetic  charge $g$.  Here we see that the free energy have a  local minimum and  a local maximum corresponding to the  extremal points of the Hawking temperature (see Fig. \ref{th0}). At these points the heat capacity flip the sign (see Fig. \ref{sh1}) .  For $r>r_{c1}$, the free energy is the increasing function of horizon radius $r$ and becomes positive at large value of $r$ and attains the maximum value at $r_{c2}$. At $r=r_{c2}$ the slope of free energy turn negative and the theory naturally provides the Hawking-Page phase transition.

{ Now we study the phase transition of the Bardeen massive black hole in the $T-S$ plane for fixed valued of massive gravity parametere. The critical values can be obtained by solving the following equation
\begin{equation}
\left(\frac{\partial T}{\partial S}\right)_{m_{g}}=\left(\frac{\partial^2 T}{\partial S^2}\right)_{m_{g}}=0
\label{ts14}
\end{equation} 
The temperature of Bardeen massive  black hole in term of entropy  is written as 
\begin{eqnarray}
T&=&\frac{1}{4\sqrt{\pi}(g^2\pi +S)}\Big(\frac{3m^2_g S^{3/2}(1+\alpha+\beta)}{\pi}+\nonumber\\&&~~\frac{cm^2_g\sqrt{\pi S}(1+2\alpha+3\beta)(\pi g^2-2S)}{\sqrt{\pi}}-\frac{1+c^2m^2_g(\alpha+3\beta)(S-2g^2\pi)}{S^{1/2}}\Big)
\label{ts13}
\end{eqnarray}

Substitution Eq. (\ref{ts13}) into Eq. (\ref{ts14}), we find the critical points  and the numerical results are presented in Tab I.
\begin{table}[ht]
 \begin{center}
 \begin{tabular}{ l | l   | l   | l    l   }
\hline
            \hline
  \multicolumn{1}{c|}{ $m_g$} &\multicolumn{1}{c}{$S_c$}  &\multicolumn{1}{|c|}{$g_c$}  &\multicolumn{1}{c}{$T_c$} \\
            \hline
            \,\,\,\,\,1~~ &~~0.768~~ & ~~0.156~~ & ~~0.017~~      \\
            \,\,\,\,\,2~~ &~~0.670~~  & ~~0.149~~ & ~~0.025~~    \\
            \,\,\,\,\,3~~ &~~0.651~~ & ~~0.148~~ & ~~0.072~~  \\
            \,\,\,\,\,4~~ &~~0.645~~ & ~~0.148~~ & ~~0.131~~  \\   
            \hline 
\hline
        \end{tabular}
        \caption{The table for critical temperature $T_c$, critical entropy $S_c$ and critical magnetic charge  $g_c$ corresponding different value of $m_g$ with fixed value of $\alpha=2,\,\beta=0.7$.}
\label{tr2}
    \end{center}
\end{table}

In order to obtained the phase transition of the black holes, we can identify the free energy of the black hole to see the effect of massive gravity parameter on the phase structure.
\begin{figure*} [h]
	\begin{tabular}{c c c c}
		\includegraphics[width=0.65\linewidth]{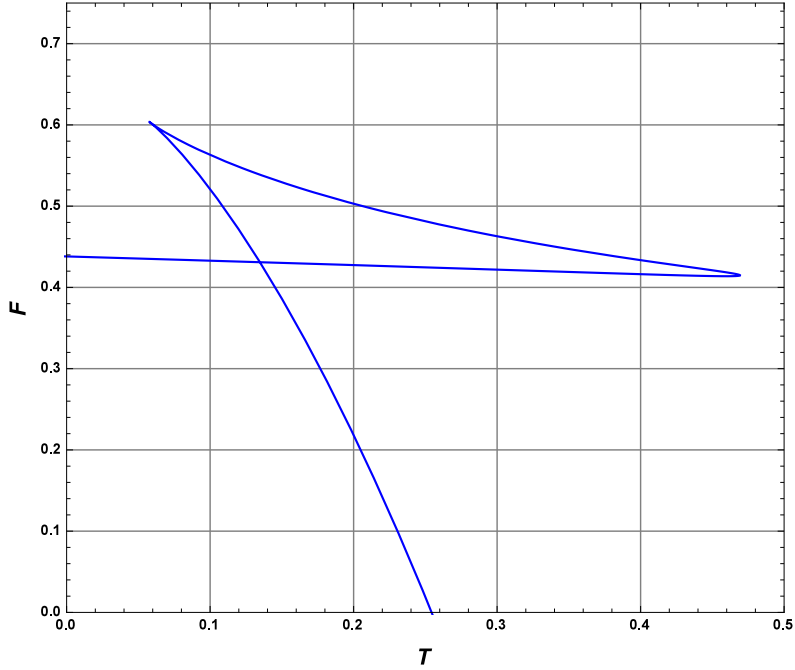}
	\end{tabular}
	\caption{ The plots of free energy vs temperature for $g<g_c$ with fixed value of $\alpha=2,\,\beta=0.7$. }
	\label{sh}
\end{figure*}

In $F-T$ plot we see that for $g<g_c$ for the Bardeen massive black hole the small and large black hole
are stable but intermediate black hole is unstable, since the heat capacity is negative (see the Fig. \ref{sh1}).  In $F\-- T$  plot the
appearance of characteristic swallow tail shows that the obtained values are critical ones in which the phase transition take place for $g<g_c$. It is worthwhile to mention that the critical value of entropy, magnetic charge decreases and temperature  increases with the massive gravity parameter (see the Tab. I).
}
%-----------------------------------------------------------------------
\subsection{Grand Canonical ensemble}
%-----------------------------------------------------------------------
Let us consider the Bardeen massive black hole in  a grand canonical ensemble,   where the black hole exchange the charge  with the surrounding and chemical potential $\phi_g$ can be held fixed. In this way, the system is being considered  in the grand canonical ensemble.  Its chemical potential $ \mu $ is specified as follows
\be
\mu=\frac{(g^2+r^2)^{3/2}(3+3r_+\gamma+r^2\Lambda+3\zeta)}{6r^2}
\ee
 The  temperature of  Bardeen massive black hole in terms of chemical potential $\mu$ is  written as 
\bea
 T=\frac{1+2r\gamma+r_+^2\Lambda+\zeta+(2+r\gamma+2\zeta)\Big(1-\sqrt{1+\frac{16 \mu^2}{(3+3r_+\gamma+r^2\Lambda+3\zeta)^2}}\Big)}{4\pi r\Big(\sqrt{2}+\sqrt{1+\frac{16 \mu^2}{(3+3r\gamma+r^2\Lambda+3\zeta)^2}}\Big)}
\eea
\begin{figure*} [h]
	\begin{tabular}{c c c c}
		\includegraphics[width=0.75\linewidth]{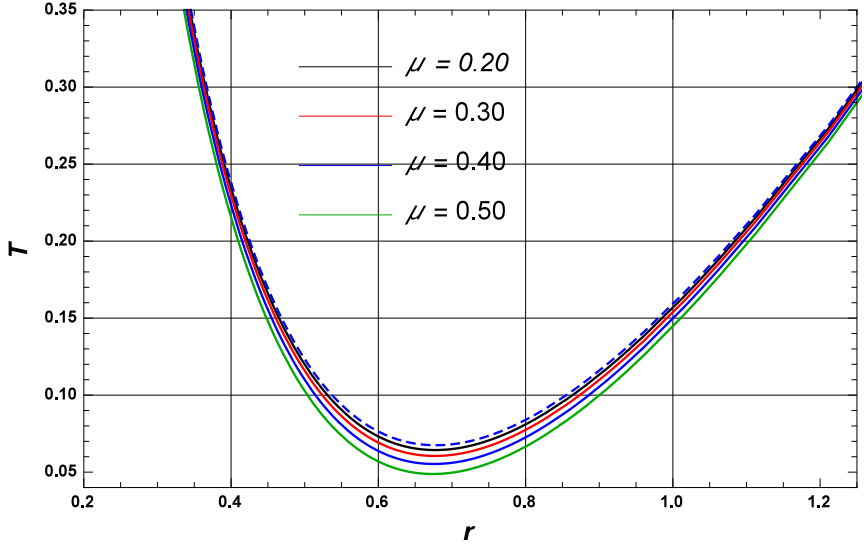}
	\end{tabular}
	\caption{ Plot of temperature $T$  as the function of horizon radius  $r$ for different value of chemical potential  $\mu$. The dotted line shows the temperature of the Schwarzschild massive black hole. }
	\label{sh7}
\end{figure*}

\begin{figure*} [h]
	\begin{tabular}{c c c c}
		\includegraphics[width=0.5\linewidth]{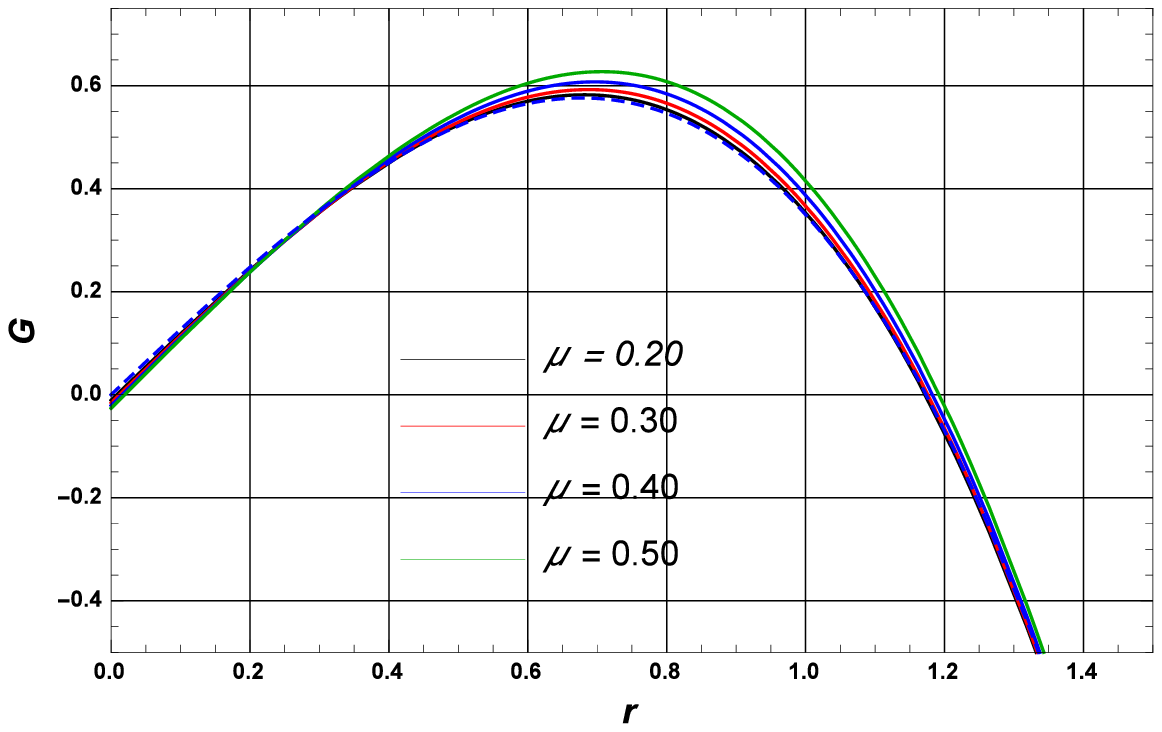}
	\includegraphics[width=0.5\linewidth]{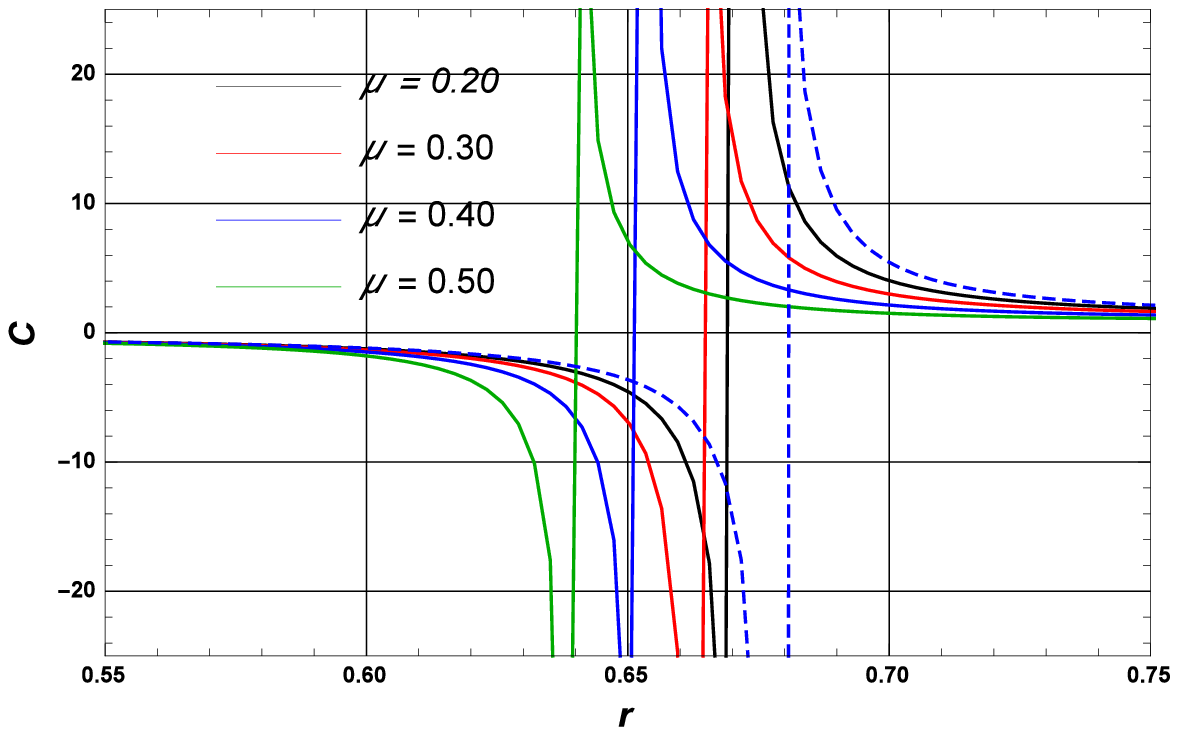}
	\end{tabular}
	\caption{ Plots of Gibbs free energy and heat capacity  vs horizon radius  $r$ for different value of chemical potential $\mu$  with $\alpha=2,\,\beta=0.7$ and $m_g^2c^2=1$. The dotted curve shows the free energy of the Bardeen black hole.  }
	\label{sh8}
\end{figure*}

In Fig. \ref{sh7} we show the variation of the temperature of  Bardeen massive black hole at different value of chemical potential $\mu$. From the Fig. \ref{sh7} it is obvious that temperature is decreasing with  increasing the horizon radius $r$ and attains the minimum value for the fixed value of chemical potential $\mu$. The minimum value of the temperature will occur at $r=r_c$ where the heat capacity diverges (see Fig. \ref{sh8} (right)). We have chosen $\alpha=2,\,\, \beta =0.7$ as a simple choice of the parameters in this region  and adopt the condition $m^2c^2=1$.

%{\bf Using the thermodynamics quantities associates with the black (mass, charge, themperature and entropy),   One can easily show that these quantities does not  follow the first law of thermodynamics.he corrected temperature is obtained from the following first law  \cite{26, Ma, Maulif}
%\be
%C_{M}dM=T\,dS + V\,dP + \mu \,dg,
%\label{mod}
%\ee}

In the case of Grand canonical ensemble the corresponding free energy is the Gibbs free energy $G$ which is expressed as $G=M-T S-\mu g $. It can be seen  from Fig. \ref{sh8} that the value of expression $  (1-\frac{\Lambda}{3}r^2+\zeta-\mu^2) $ will decide the sign of free energy  that is globally thermodynamic stability exists when the following condition is satisfied.
\be
\Lambda r^2\geq 3\left(1+\zeta-\mu^2\right)
\ee

In addition locally thermodynamic stability can also be verified by analysing the heat capacity $C$, which is shown in Fig. (\ref{sh8}). The condition of locally thermodynamic stability is given by $\Lambda r^2>\left(1+\zeta-\mu^2\right)$. In this grand canonical aspect both  global  and local stability depend on the parameters $ \Lambda,c_1 $ and also on $ \mu $ unlike canonical case where stability conditions depends only upon $ \Lambda $ and $ \zeta $. Once again both these parameters depend upon $ \alpha $ and $ \beta $.

%{\bf From the expression of temperature and entropy the relation between the temperature and entropy for the dRGT massive Bardeen black hole is
%\begin{eqnarray}
%T&=&
%\label{tsg}
%\end{eqnarray} 

\begin{figure*} [h]
	\begin{tabular}{c c c c}
		\includegraphics[width=0.65\linewidth]{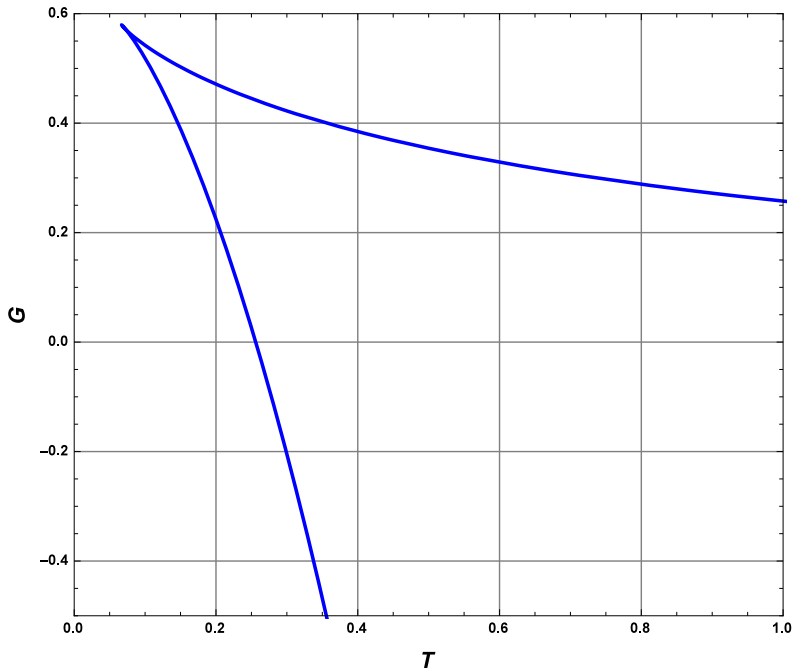}
	\end{tabular}
	\caption{ The plots of  Gibbs free energy vs. temperature for  $\mu<\mu_c$ with fixed value of $\alpha=2,\,\beta=0.7$ and $\mu_c=1.028$. }
	\label{sh9}
\end{figure*}

 The plot of Gibbs free energy  with the function of temperature is plotted in the Fig \ref{sh9}. In this figure we notice that the Bardeen massive black hole is stable above the Hawking temperature, where the Gibbs free energy is negative.

%\begin{table}[ht]
% \begin{center}
% \begin{tabular}{ l | l   | l   | l    l   }
%\hline
%            \hline
%  \multicolumn{1}{c|}{ $m$} &\multicolumn{1}{c}{$S_C$}  &\multicolumn{1}{|c|}{$\mu_C$}  &\multicolumn{1}{c}{$T_C$} \\
%            \hline
%
%            \,\,\,\,\,1~~ &~~0.768~~ & ~~0.159~~ & ~~0.017~~      \\
%            %
%            \,\,\,\,\,2~~ &~~0.670~~  & ~~0.149~~ & ~~0.025~~    \\
%            %
%            \,\,\,\,\,3~~ &~~0.651~~ & ~~0.148~~ & ~~0.072~~  \\
%                %
%            \,\,\,\,\,4~~ &~~0.645~~ & ~~0.148~~ & ~~0.131~~  \\   
%            \hline 
%            %
% 
%\hline
%        \end{tabular}
%        \caption{The table for critical temperature $T_C$, critical pressure $S_C$ and $g_c$ corresponding different value of $m$ with fixed value of  $\alpha=2,\,\beta=0.7$ and $m_g^2c^2=1$  .}
%\label{tr2}
%    \end{center}
%\end{table}

For  further analysis of the stability of the black hole as shown in Fig. \ref{sh8} for the different values of chemical potential $\mu$,  it is shown that  heat capacity  for a given value of  chemical potential $\mu$ is discontinuous exactly at the critical radius $r_c$. Further, we note that there is a flip of sign in the heat capacity around $r_c$ . Thus  Bardeen massive black hole is   thermodynamically stable for  $r > r_c$ whereas it is thermodynamically unstable for $r<r_c$ and there is a phase transition at $r=r_c$ from the unstable to stable phase. 
%-----------------------------------------------------------------------
\section{Conclusion}
%-----------------------------------------------------------------------
 The
dRGT massive gravity describes non-linear interaction terms as a correction of the Einstein-Hilbert action and reduces to general relativity as a particular limit. The  dRGT massive gravity has received significant attention including searches for the solutions of black holes. We note that because of the inclusion of the massive gravity term in the action,  the Bardeen black hole solution  is modified and the corresponding thermodynamic quantities are also changed. The temperature decreases with increasing the horizon radius and attains a minimum value and negative at the larger value of magnetic charge. The temperature of the Bardeen massive  black hole coincide with the Schwarzschild massive black hole at $r=1.1$ for canonical ensemble and $r=1.4$ for grand canonical ensemble. However, the black hole entropy do not obey the area law, a new quantity  $C_M$ is defined which is required for the consistency of the first law of thermodynamics and area law. We studied the black hole thermodynamics in both canonical and grand canonical ensembles to analyse the thermodynamic quantities including the phase transition. The stability of the black hole has been studied by  observing the behaviour of heat capacity and the free energy. The slope of free energy becomes negative after the heat capacity diverges and the Hawking-Page phase transition occurs naturally in both the ensembles.

 The free energy turns  negative slope where the  heat capacity is diverges and the Hawking-Page phase transition occur in both ensembles.

In this study we constructed the  exact  solution of  Bardeen  black holes in the presence  dRGT massive gravity which  reduces to   Bardeen black holes when $m_g=0$ and $AdS$ massive black hole in the absence of charge.  The resulting black hole solution is  characterized by analysing horizons, which at most could be threefold , so that  inner, outer and  cosmological horizons. We have also analysed the thermodynamic quantities like the black hole mass, Hawking temperature, entropy and free energy at event horizon  in terms of magnetic charge $g$ and the massive gravity parameter $m_g$ in both canonical and grand canonical ensemble. The thermodynamics  of the black hole is modified due to the  presence of non-linear source. The local and global stability of the black hole  fot the case of grand canonical ensemble have been studied by investigating the  heat capacity and Gibbs free energy. The heat capacity flip the sign at $r=r_c$, where the temperature is minimum. The positive heat capacity $C>0 $ for $r<r_c$ allowing the black hole to become thermodynamically stable and the  black holes are globally preferred with negative free energy.

We further analysed the stability of  canonical ensemble  by studying the nature of Helmoltz free energy.  Free energy has a local minima and local maxima corresponding to the horizon radius where the specific heat diverges (see Fig.4 b) and these points can be  identified as the extremal points of the Hawking temperature (see Fig.\ref{th0}).  However, at very small horizon radius the Hawking temperature is  negative and hence not physical for global stability.  This is  in accordance with the Hawking-Page phase transition in general relativity \cite{li}.
\section*{Acknowledgement}
Authors  would like to thank Dr. Dharm Veer Singh for useful discussions.
%%%%%%%%%%%%%%%%%%%%%%%%%%%%%%%%%%%%%%%%%%%%%%%%%%%%%%%%%%

\end{document}